\documentclass[draft,11pt,english]{article}
\usepackage{amsfonts,amsmath,amsthm,amscd,amssymb,latexsym}

\newcommand{\supp}{\operatorname{supp}}
\newcommand{\const}{\operatorname{const}}

\numberwithin{equation}{section}

\newtheorem{theorem}{Theorem}

\begin{document}

\title{
Global in Time Madelung Transformation for Kolmogorov-Feller
Pseudodifferential Equations\thanks{This work was supported by DFG
project 436 RUS 13/895/0-1.}}

\author{S.~Albeverio$^1$ and V.~G.~Danilov$^2$\\
\\
$^1${\small Inst. Appl. Mathematics, SFB611, IZKS, University of Bonn,}\\
{\small BiBoS, Universities of Bielefeld and Bonn,}\\
{\small CERFIM (Locarno) Acc. Arch., Usi (Mendrisio)}\\
\\
$^2${\small Moscow Technical University of Communications and Informatics,}\\
{\small Moscow Institute of Electronics and Mathematics}
}


\date{}

\maketitle

\begin{abstract}
Using an idea going back to Madelung we construct global in time
solutions to the transport equation corresponding to the asymptotic
solution of the Kolmogorov-Feller equation describing a system with
diffusion, potential and jump terms. To do that we use the
construction of a generalized delta -shock solution of the
continuity equation for a discontinuous velocity field. We also
discuss corresponding problem of asymptotic solution construction
(Maslov tunnel asymptotics).
\smallskip

Kolmogorov--Feller equation,
diffusion, jump process, delta-shock solutions.
\smallskip

\end{abstract}

\section*{Introduction}

The goal of the present paper is to present a new approach to the
construction of singular (i.e., containing the Dirac
$\delta$-function as a summand) solutions to the continuity equation
and to show how these solutions can be used to construct the global
in time solution of the Cauchy problem for Kolmogorov--Feller-type
equations with diffusion, potential and jump terms. The relation
between the solutions of the continuity equation and the system
consisting of the Hamilton--Jacobi equation plus the transport
equation has been well studied before in the case of a smooth action
functional.

The velocity field $u$ is determined as the family of velocities of
points on the projections of the trajectories of the Hamiltonian
system corresponding to the Hamilton--Jacobi equation.  As it was
mentioned by E. Madelung [13], in this velocity field, the squared
solution of the transport equation (denoted by $/rho$) satisfies the
continuity equation
\begin{equation}\label{0.1}
\rho_t + (\nabla, u\rho)+a\rho=0
\end{equation}
with some additional term $a\rho$, which is defined below ($a$ is
equal to 0 if the Hamiltonian is formally self-adjoint). The main
obstacle to the extension of this correspondence globally in time is
the fact that in general the solution of the Hamilton-Jacobi
equation  are smooth only locally in time. The loss of smoothness is
equivalent to the appearance of singularities of the velocity field
mentioned above. Till recent times there was no method for
constructing formulas for solutions of the continuity equation for a
discontinuous velocity field. It is clear that the continuity
equation has the divergent form and this very important property
allows precisely to introduce the concept of a global solution in
spite of singularities in the velocity field. The divergent form as
itself does not play any role in Madelung's approach but it is very
important for our construction because we deal with singular
solutions.

In the present paper we generalize Madelung's approach to the case
in which the singular support of the velocity field is a stratified
manifold transversal to the velocity field trajectories. This holds,
for example, in the case where the space is one-dimensional  under
the condition that, for any $t\in[0,T]$, the velocity field singular
support is a discrete set without limit points.

It is of interest to note that the generalization of Madelung's idea, first suggested for the
Schr\"odinger equation, also makes sense for equations from a different class.

Namely, let $\Lambda^n$ be the Lagrangian manifold corresponding to the solution and let $\pi\colon
\Lambda^n\to \mathbb{R}^n_x$ be the projection mapping.

All the points of $\pi^{-1}(x)\in\Lambda^n$ make contributions to the construction of an asymptotic
oscillating solution (of the type of the WKB-solution $\varphi(x,t)\exp(iS(x,t)/h)$).

But if we consider nonoscillating solutions of the form $\varphi(x,t)\exp(-s(X,T)/H)$, just as for
equations of heat conduction type with a small parameter $h$ at the second-order derivative, see
Sec.~2 and [8], then only the point from the set $\bar{\gamma}\in\pi^{-1}(x)$ at which
$$
S(\bar{gamma},t)=\min_{\bar{gamma}\in \pi^{-1}(x)}S(\gamma,t)
$$
makes a contribution to the solution. hence the points of the Lagrangian manifold making
contributions to the construction of a nonoscillating solution form a surface with (shock wave-type)
jumps, which results in discontinuities of $\nabla S$ and hence in discontinuities of the velocity
field for the transport equation.

The last equation does not have a divergent form in contrast to the continuity equation, which
arises in the Madelung construction. So Madelung's idea was initially adopted to the global
constructions involving generalized solutions to the continuity equation.

In Section~2, we describe the definition and construction of these types of solutions. The
conclusion is that these solutions can be constructed by means of characteristics. This allows us
partly to change the direction of time and to solve some inverse problems. We will discuss this
problem in detail in subsequent papers.

\section{Generalized solutions\\ of the continuity equation}

Here we follow the approach developed in [1], where the solution of
the continuity equation is understood in the sense of an integral
identity, which, in turn, follows from the fact that relation
(\ref{0.1}) can be understood in the sense of the distributional
space $\mathcal{D}'(\mathbb{R}^{n+1}_{x,t})$. The first step in this
way has been done in [14], see also [15], [16], where the approach
based on smooth approximations of the solutions was used.

We specially note that the integral identities in [1] can be derived
without using the construction of nonconservative products [2, 4] of
the nonsmooth and generalized functions (or measure solutions [5]),
and the value of the velocity on the discontinuity lines (surfaces)
is not given a priori but is calculated. In the case considered in
[1], the integral identities exactly coincide in form with the
identities derived using the construction of a nonconservative
product (measure solutions) in the situation described at the end of
above introduction, which we shall now make more precise.

First, we consider an $n-1$-dimensional surface
$\gamma_t$ moving in $\mathbb{R}^n_x$,
which is determined by the equation
$$
\gamma_t=\{x;t=\psi(x)\},
$$
where $\psi\in C^1(\mathbb{R}^n)$, and $\nabla\psi\ne0$ in the
domain in $\mathbb{R}^n_x$ where we work.

This is equivalent to determining a surface
by an equation of the form
$$
S(x,t)=0
$$
($S\in C^1$ in both variables, $S(x,t)=0$, $\nabla_{x,t}
S|_{S=0}\ne0$) under the condition that
$$
\frac{\partial S}{\partial t}\ne0.
$$

We remain that the situation with $\frac{\partial S}{\partial t}=0$,
can also be covered by making the change of variables
$x'_i=x_i-c_it$ with appropriately chosen $c_i$, $i=1...n$, solving
the problem with the moving surface  and then returning to the
original variables. Possible generalizations are considered later in
this section.

Next, we assume that $\zeta$ belongs to
$C^\infty_0(\mathbb{R}^n\times\mathbb{R}^1_+)$. Then, by definition,
$$
\langle\delta(t-\psi(x)),\zeta(x,t)\rangle
=\int_{\mathbb{R}^n}\zeta(x,\psi(x))\,dx,
$$
where $\delta$ is the Dirac delta function and $\langle,\rangle$ is
the distributional pairing (with respect to the variable
$t\in\mathbb{R}^1_+$ and $x\in\mathbb{R}^n$).

Let $\delta(t-\psi(x))$ be applied to the test function $\eta\in
C^\infty_0(\mathbb{R}^n)$, then
$$
\langle\delta(t-\psi(x)),\eta(x)\rangle =\int_{\gamma_t}
\eta\,d\omega_\psi,
$$
where $d\omega$ is the Leray form [6] on the surface
$\{t=\psi(x)\}$
such that $-d\psi d\omega_\psi=dx_1\dots dx_n$.

One can show that (see [1], [6])
$$
\langle\delta(t-\psi(x)),\eta(x)\rangle =\int_{\gamma_t}
\frac{\eta(x)}{|\nabla\psi|}\,d\sigma.
$$

First, we assume that  the solution $\rho$
to Eq.~(\ref{0.1}) has the form
\begin{equation}\label{2.1}
\rho(x,t)=R(x,t)+e(x)\delta(t-\psi(x)),
\end{equation}
where $R(x,t)$ is a piecewise smooth function
with possible discontinuity at $\{t=\psi(x)\}$:
$$
R=R_0(x,t)+H(t-\psi(x))R_1(x,t),
$$
$e\in C(\mathbb{R}^n)$ and has a compact support, $\psi\in C^2$ and
$\nabla \psi\ne0$ for $x\in\supp e$, and $H(z)$ is the Heaviside
function.

It is clear that the term
$$
e(x)\delta'(t-\psi(x))
$$
appears in (\ref{0.1}) if we differentiate the distribution
$\delta(t-\psi(x))$ with respect to $t$. Hence it is necessary to
have in (\ref{0.1})
$$
( \nabla,\rho u)=-e(x)\delta'(t-\psi) +  \text {smoother summands},
$$
since $\nabla\delta(t-\psi)=-\nabla\psi\delta'(t-\psi)$.
Then we must have
$$
\rho u=\frac{e\nabla\psi}{|\nabla\psi|^2}\delta(t-\psi) +\text{
smoother summands}.
$$

Now we formulate an integral identity, defining a generalized
solution to the continuity equation.

We set $\Gamma=\{(x,t); t=\psi(x)\}$; this is an $n$-dimensional
surface in $\mathbb{R}^n\times\mathbb{R}^1_+$. Let
$$
u(x,t)=u_0(x,t)+H(t-\psi)u_1(x,t),
$$
where $\psi$ is the same function as before, and $u_0,u_1\in C
(\mathbb{R}^n\times \mathbb{R}^1_+)$.

Let us consider Eq.~(\ref{0.1}) in the sense of distributions.
For all $\zeta(x,t)\in C^\infty_0(\mathbb{R}^n\times\mathbb{R}^1_+)$,
$\zeta(x,0)=0$, we have
$$
\Big\langle\frac{\partial \rho}{\partial t} +(\nabla,\rho
u),\zeta\Big\rangle =-\langle\rho,\zeta_t\rangle-\langle\rho
u,\nabla\zeta\rangle.
$$
Substituting the singular terms for $\rho$ and $\rho u$ calculated above,
we come to the following definition.
\medskip

{\bf Definition 1.1}
A function
$\rho(x,t)$ determined by relation (\ref{2.1}) is called a generalized
$\delta$-shock wave type solution to (\ref{0.1})
on the surface $\{t=\psi(x)\}$ if the integral identity holds
\begin{align}\label{1.4}
&\int^\infty_0\int_{\mathbb{R}^n}
(R\zeta_t+(uR,\nabla \zeta)+aR\zeta)\,dx\,dt
\nonumber\\
&\qquad
+\int_{\Gamma}\frac{e}{|\nabla\psi|}\frac{d}{dn_\perp}\zeta(x,t)\,dx=0
\end{align}
for all test functions $\zeta(x,t)\in C^\infty_0(\mathbb{R}^n\times
\mathbb{R}^1_+)$, $\zeta (x,0)=0$,
$\frac{d}{dn_\perp}=\big(\frac{\nabla\psi}{|\nabla\psi|},\nabla\big)
+|\nabla\psi|\frac{\partial}{\partial t}$.
\medskip

We have also the relation
$$
\int_{\mathbb{R}^n}\frac{e}{|\nabla\psi|}\frac{d}{dn}\zeta(x,\psi)\,dx
=\int_{\Gamma}\frac{e}{|\nabla\psi|}\frac{d}{dn_\perp}\zeta(x,t)\,dx.
$$

We note that the vector $n_\perp$ is orthogonal to the vector
$(\nabla\psi,-1)$, which is the normal to the surface $\Gamma$,
i.e., $\frac{d}{dn_\perp}$ lies in the plane tangent to $\Gamma$.

We can give a geometric definition of the field $\frac{d}{dn_\perp}$.
The trajectories of this vector field are curves lying on the surface
$\Gamma$, and they are orthogonal to all sections of this surface
produced by the planes $t=\operatorname{const}$.
Furthermore, it is clear that
the expression $\frac1{|\nabla\psi|}$ is an absolute value
of the normal velocity of a point on $\gamma_t$, i.e.,
on the cross-section of $\Gamma$ by the plane
$t=\operatorname{const}$,
and the expression
$\frac{1}{|\nabla\psi|}\cdot\frac{\nabla\psi}{|\nabla\psi|}
\overset{\text{def}}{=}\vec V_n$
is the vector of normal velocity of a point on $\gamma_t$.
Thus,
we have another representation:
$$
\int_{\Gamma}\frac{e}{|\nabla\psi|}\frac{d}{dn_\perp}\zeta(x,t)\,dx
=\int_{\Gamma}e\Big((\vec{V}_n,\nabla)+\frac{\partial}{\partial t}\Big)\zeta(x,t)\,dx,
$$
where $V_n=\pi^*(v_n)$, $v_n$ is the normal velocity of a point on
$\gamma_t$, and $\pi^*$ is induced by the projection mapping
$\pi:\Gamma\to R^n_x $.

It follows from the latter definition that the following relations
must hold:
\begin{equation}\label{1.12}
R_t+(\nabla, Ru)+aR\zeta=0, \qquad\text{for all points}\quad (x,t)\not\in\Gamma,
$$
$$
([R]-|\nabla\psi|[Ru_n])+\bigg(\frac{d}{dn}\bigg)^*\frac{e}{|\nabla\psi|}=0,\qquad
\text{for all points}\quad(x,t)\in\Gamma,
\end{equation}

The last relation can be rewritten in the form
\begin{equation}\label{1.3}
\mathcal{K} E+\frac{d}{dn} E =[Ru_n]|\nabla\psi|-[R],
\end{equation}
where $E=e/|\nabla\psi|$, the factor
$\mathcal{K}=(\nabla,\frac{\nabla\psi}{|\nabla\psi|})=\operatorname{div}
\nu$ ($\nu$ is the normal on the surface $\{t=\psi(x)\}$) and, as is
known, is the mean curvature of the cross-section of the surface
$\Gamma$ by the plane $t=\const$,
$\frac{d}{dn}=(\frac{\nabla\psi}{|\nabla\psi|},\nabla)$.

Now we assume that there are two surfaces
$$
\Gamma_i=\{(x,t); t=\psi_i(x)\}
$$
in $\mathbb{R}^n\times\mathbb{R}^1_+$, $i=1,2$,
whose intersection is a smooth surface
$$
\hat\gamma=\{(x,t);(t=\psi_1)\cap(t=\psi_2)\}
$$
belonging to the third surface
$\Gamma_{(3)}=\{(x,t);t=\psi_3(x)\}$. Further, we assume that the
surface $\Gamma_{(3)}$ is a continuation of the surfaces
$\Gamma^{(i)}$ in the following sense.
We let $n^{(i)}_\perp$ denote curves on the surfaces $\Gamma_i$
and we assume that each point $(\hat{x},\hat{t})$ on the surface  $\hat{\gamma}$
is associated with a graph consisting of the trajectories $n^{(1)}_\perp$ and
$n^{(2)}_\perp$ entering $(\hat{x},\hat{t})$ and the trajectory
$n^{(3)}_\perp$ leaving this point (i.e., the trajectories
$n^{(i)}_\perp$ fiber the surface $\Gamma^{(i)}$).
We also assume
that the surface (stratified manifold)
$\Gamma_\cup=\Gamma_{(1)}\cup\Gamma_{(2)}\cup\Gamma_{(3)}$ consists
of points belonging to these graphs. Next, we assume that $u(x,t)$
is a piecewise smooth vector field whose trajectories enter
$\Gamma_\cup$.

{\bf Definition 1.2.} Let
$$
u(x,t)=u_0(x,t)+\sum^3_{i=1}H(t-\psi_i)u_{1i}(x,t),
$$
where $\psi$ is the same function as before, and $u_0,u_{1i}\in
C(\mathbb{R}^n\times \mathbb{R}^1_+)$. The function $\rho(x,t)$
determined by the relation
$$
\rho(x,t)=R(x,t)+\sum^{3}_{i=1}e_i(x)\delta(t-\psi_i(x)),
$$
where $R(x,t)\in C^1(\mathbb{R}^n\times\mathbb{R}^1_+)
\setminus\{\bigcup\Gamma^{(i)}_t\}$,
is called a generalized $\delta$-shock wave type solution
to (\ref{1.4}) corresponding to the stratified manifold
$\Gamma_\cup$ if the integral identity
\begin{align}\label{2.4}
&\int^\infty_0\int_{\mathbb{R}^n} (R\zeta_t+(uR,\nabla
\zeta)+aR\zeta)\,dx\,dt
\nonumber\\
&\qquad +\sum^3_{i=1}
\int_{\Gamma_{(i)}}\frac{e_i}{|\nabla\psi_i|}\frac{d}{dn^{(i)}_\perp}\zeta(x,t)\,dx=0
\end{align}
holds for all test functions
$\zeta(x,t)\in C^\infty_0(\mathbb{R}^n\times
\mathbb{R}^1_+)$, $\zeta(x,0)=0$,
$\frac{d}{dn^{(i)}_\perp}=\big(\frac{\nabla\psi_i}{|\nabla\psi_i|},\nabla\big)
+|\nabla\psi_i|\frac{\partial}{\partial t}$.

As above this relation implies the first equation from (\ref{1.12})
outside $\Gamma_\cup$, equations of the type of the second equation in (\ref{1.12})
on strata $\Gamma_{(i)}$ and the Kirchhoff type relation on $\hat\gamma$:
\begin{equation}\label{1.13}
(e_1+e_2)\big|_{\hat\gamma} =e_{3} \big|_{\hat\gamma}.
\end{equation}

Now we consider the case with $\operatorname{codim} \Gamma>1$.
First,
we note that the second integral in (1.2) can be written as
$$
\int_{\Gamma
}\frac{e}{|\nabla\psi|}\frac{d}{dn_\perp}\zeta(x,t)\,dx
=\int_{\Gamma}e\bigg(\bigg(\frac{\nabla\psi}{|\nabla\psi|^2},\nabla\bigg)
+\frac{\partial}{\partial t}\bigg)\zeta(x,t)\,dx.
$$
We note that if the surface $\Gamma$ is determined by the equation
$S(x,t)=0$ rather than by the simpler equation $\{t=\psi(x)\}$
presented at the beginning of this section, then
$$
\vec V_n=-\frac{S_t}{|\nabla S|}\cdot \frac{\nabla S}{|\nabla S|}
=-\frac{S_t}{|\nabla S|^2}\nabla S
$$
and, of course, the new vector field $\frac{d}{dn_\perp}=(\vec
V_n,\nabla) +\frac{\partial }{\partial t}$ remains tangent to
$\Gamma$.

Therefore, in this more general case, using this new vector $\vec V_n$,
we can again rewrite the integral identity from Definition~1.1 as
\begin{equation}\label{2.5}
\int^\infty_0\int_{\mathbb{R}^n}
\big(R\zeta_t+(uR,\nabla\zeta)+aR\zeta\big)\,dx dt +
\int_{\Gamma}e\Big((\vec V_n,\nabla)
+\frac{\partial}{\partial t}\Big)\zeta(x,t)\,dx =0.
\end{equation}

This form of integral identity can easily be generalized to the case
in which $\Gamma$ is a smooth surface in $\mathbb{R}^{n+1}$ of
codimension~$>1$.

In this case, instead of $\vec V_n$, we can use a vector $\vec v$
that is transversal to $\Gamma$ and such that the field $(\vec
v,\nabla)+\frac{\partial}{\partial t}$ is tangent to $\Gamma$. We
note that the vector $\vec v$ is uniquely determined by this
condition,  which can be treated as ``the calculation of the
velocity value on the discontinuity'' from the viewpoint of [5] and
[7].

Moreover, in this case, the expression for $\rho$ does not contain
the Heaviside function, and it is assumed that the trajectories of
the field $u$ are smooth, nonsingular outside $\Gamma$, and
transversal to $\Gamma$ at each point of $\Gamma$.
In this case,
the function $\rho$ has the form
$$
\rho(x,t)=R(x,t)+e(x)\delta(\Gamma),
$$
where $R\in C^1(\mathbb{R}^{n+1}\setminus\Gamma)$, $e\in
C^1(\Gamma)$, and the  function $\delta(\Gamma)$ is determined
by
$$
\langle\delta(\Gamma),\zeta(x,t)\rangle=\int_{\Gamma}\zeta\omega,
$$
where $\omega$ is the Leray form on $\Gamma$. If
$\Gamma=\{S_1(x,t)=0\cap\dots\cap S_k(x,t)=0\}$, $k\in[1,n]$, then
$\omega$ is determined by the relation, see [6], p.~274,
$$
dt\,dx_1\cdot\dots\cdot d\,x_n = dS_1\cdot\dots\cdot dS_k\omega.
$$

In this case, we assume that the functions $S_k$ are sufficiently
smooth (for example, $C^2(\mathbb{R}^n\times\mathbb{R}^1_+)$) and
their differentials on $\Gamma$ are linearly independent.

Moreover, we can assume that the inequality
$$
J=\frac{\mathcal{D}(S_1,\dots,S_n)}{\mathcal{D}(t,x_1,\dots,x_{n-1})}\ne0
$$
holds. This inequality is an analog of $S_t\ne0$ at the beginning of
this section and allows us to write $\omega$ in the form
$$
\omega=J^{-1}dx_k\cdot\dots\cdot dx_n.
$$

The integral identity, an analog of (\ref{2.5}), has the form
$$
\int^\infty_0\int_{\mathbb{R}^n}
\big(R\zeta_t+(uR,\nabla\zeta)+aR\zeta\big)\,dx dt +
\int_{\Gamma}e\bigg((v,\nabla)+\frac{\partial}{\partial t}\bigg)
\zeta(x,t)\omega =0.
$$
Integrating the latter relation by parts, we obtain equations for
determining the functions $e$ and $R$ similarly to (\ref{1.3}).

Now we assume that the singular support of the velocity field
is the stratified manifold $\bigcup \Gamma_{i}$
with smooth strata $\Gamma_{t}$ of codimensions $n_i\geq1$.

We also assume that the velocity field trajectories
are transversal to $\bigcup\Gamma$ and are entering trajectories.

Then the general solution of  Eq.~(\ref{0.1}) has the form
$$
\rho(x,t)=R(x,t)+\sum e_i \delta(\Gamma_i),
$$
where $R(x,t)$ is a function smooth outside $\bigcup \Gamma_{i}$,
$e_i$ are functions defined on the strata $\Gamma_{i}$, and the sum
is taken over all strata $\Gamma_{i}$.

The integral identities determining such a generalized solution
have the form
\begin{align} \label{2.6}
&\int^\infty_0\!\int_{\mathbb{R}^n}
(R\zeta_t+(uR,\nabla\zeta)+aR\zeta)\,dx dt
\nonumber\\
&\qquad
+\sum_{i}\int_{\Gamma_{i}} e_i
\bigg[\bigg((v_i,\nabla)+\frac{\partial}{\partial t}\bigg)\zeta(x,t)\bigg]\omega_i=0.
\end{align}
This implies that, outside $\bigcup \Gamma_{i}$,
the function $R$ satisfies the continuity equation
$$
R_t + (\nabla, uR)+aR=0,
$$
and, on the strata $\Gamma_{j}$ for $n_j=1$,
equations of the form (\ref{1.3}) hold, which contain the values of $R$
brought to $\Gamma_{t}$ along the trajectories.
For $n_l=n-k$, $k>1$, on the strata $\Gamma_{l}$,
we have the equations
\begin{equation}\label{2.7}
\frac{\partial}{\partial t} e_l\mu_l +(\nabla,v_le_l\mu_l)=F_l\mu_l,
\end{equation}
where $\mu_l$ is the density of the measure $\omega_l$ with respect
to the measure on $\Gamma_{l}$ which is left-invariant with respect
to the field $\frac{\partial}{\partial t}+\langle
v_l,\nabla\rangle$, and $F_l$ is defined by the following
construction. Denote a $\varepsilon$-neighborhood of $\Gamma_{l}$
by $\Gamma_{l}^\varepsilon$ and denote its boundary by
${\partial}\Gamma_{l}^\varepsilon$. Let us consider the integral
appearing after integration by parts:
$$
\int_{{\partial}\Gamma_{l}^\varepsilon}\zeta\rho u_{nl}
 \omega^{\varepsilon}_l,
$$
where  $u_{nl}$ is the normal component of the velocity $u$ on
${\partial}\Gamma_{l}^\varepsilon$,  $\omega^{\varepsilon}_l$ is
the Leray  measure on ${\partial}\Gamma_{l}^\varepsilon$, $\zeta$
is a test function. Passing to the limit as $\varepsilon\to 0$ we
obtain
$$
\lim_{\varepsilon\to0}\int_{{\partial}\Gamma_{l}^\varepsilon}\zeta\rho
u_{nl}\omega^{\varepsilon}_l =\int_{\Gamma_{l}}\zeta\ F_{l}
\omega_l.
$$

It is well known that outside $\bigcup\Gamma_{i}$ the function
$R(x,t)$ can be calculated using the famous Cauchy formula
\begin{equation}\label{3.8}
R(x.t)=\rho_{0}(x,t)\big|\frac
{Dx}{Dx_{0}}\big|^{-1}\exp(-\int^t_0adt')
\end{equation}
where $\rho_{0}$ ia a constant along the trajectories of the field u
outside  $\bigcup\Gamma_{i}$, $\big|\frac {Dx}{Dx_{0}}\big|$ is the
jacobian of the mapping corresponding to the shift along the
trajectories of $u$ and the integral under exponent is calculating
along the trajectories of the field $u$.

This formula implies that the limit as $\varepsilon\to 0$ of the
above integral exists.

 We note that it follows from the above that the
function $R$ is determined independently of the values of $v_i$ on
the strata under the condition that the field trajectories enter
$\bigcup\Gamma_{i}$.

In conclusion, we consider the case where the coefficient $a$ has
a singular support on $\bigcup\Gamma_{i}$, i.e.,
$$
a=f(u).
$$
In this case, we set
$$
a\rho=\check{a}\rho+\sum f(v_i) e_i \delta(\Gamma_{i}).
$$
where $\check{a}=f(u)$ outside $\bigcup\Gamma_{i}$. We note that
such a choice of the definition of the term $a\rho$ is not unique in
this case. But, first, it is consistent with the common concept of
measure solutions (see [3],[5]) and, second, it is of no importance
for the construction of the solution outside $\bigcup\Gamma_{i}$
for the case in which the trajectories $u$ enter
$\bigcup\Gamma_{i}$.

In this case, identity (\ref{2.6}) takes the form
\begin{align}\label{2.11}
&\int^\infty_0\int_{\mathbb{R}^n}
\big(R\zeta_t+( uR,\nabla\zeta)+f(u)R\zeta\big)\,dx dt
\nonumber\\
&\qquad
+\sum_{i}\int_{\Gamma_{i}}e_i\Big[
\Big((v_i,\nabla)+\frac{\partial}{\partial t}+f(v_i)\Big)\zeta(x,t)
\Big]\omega_i=0,
\end{align}
and Eq.~(1.7) can be rewritten in the form
\begin{equation}\label{2.12}
\frac{\partial}{\partial t}(e_l\mu_l) +( \nabla,v_le_l\mu_l)
+f(v_l)=F_l\mu_l.
\end{equation}

All the afore said gives the following statement.
\begin{theorem}
Let that the following conditions be satisfied
for $t\in[0,T]$, $T>0$:

{\rm(1)} $\bigcup\Gamma_{i}$ is a stratifies manifold with smooth strata $\Gamma_{i}$;

{\rm(2)} the trajectories of the field $u$ are smooth outside $\bigcup\Gamma_{i}$,
enter $\bigcup\Gamma_{i}$ and do not intersect outside $\bigcup\Gamma_{i}$;

{\rm(3)} equations (\ref{2.12}) are solvable on the strata $\Gamma_{i}$;

{\rm(4)} the Kirchhoff laws are satisfied on the intersections of strata $\Gamma_{i}$.

Then there exist a general solution to the continuity equation~(0.1)
with $a=f(u)$ in the sense
of the integral identity (\ref{2.11}).
\end{theorem}

\section{The Maslov tunnel asymptotics}

We recall that the asymptotic solutions of a general Cauchy problem
for an equation with pure imaginary characteristics was first
constructed by V.~P.~Maslov [8]. In the present paper, we consider
only the following Cauchy problem
\begin{equation}\label{t.1}
-h\frac{\partial u}{\partial t}
+P\bigg(\overset{2}{x},-h\overset{1}{\frac{\partial}{\partial
x}}\bigg)u=0, \qquad u(x,t,h)|_{t=0}=e^{-S_0(x)/h}\varphi^0(x),
\end{equation}
where $P(x,\xi)$ is the (smooth) symbol of the Kolmogorov--Feller
operator [9], $S_0\geq0$ is a smooth function, $\varphi^0\in
C^\infty_0$, $ h\to+0$ is a small parameter characterizing the
frequency and the amplitude of jumps of the Markov stochastic
process having having transition probability given by $P(x,\xi)$.
To be more precise, we can have in the mind the following form of $P(x,\xi)$:
$$
P(x,\xi)=(A(x)\xi,\xi) +V(x)+
\int_{\mathbb{R}^n}\bigg(e^{i (\xi,\nu)}-1\bigg)\mu(x,d\nu),
$$
where $A(x)$ is positive definite smooth matrix,
$V(x)$ is a smooth function and its second derivatives
are assumed to be uniformly bounded,
and $\mu(x,d\nu)$ is a family of bounded measures
smooth with respect to $x$ in the sense that the functions
$x\mapsto\mu(x,B)$ are smooth for all measurable sets $B$.
The symbol $P(x,\xi)$ can also depend on $t$, we will be more precise
later on.

Locally in $t$, an asymptotic solution of problem (\ref{t.1}) can be
constructed according to the scheme of the WKB method, see [8]: the
solution is constructed in the form
$$
u=e^{-S(x,t)/h}\sum_{i=0}^{\infty}(\varphi_i(x,t)h^{i}
$$
in the sense of asymptotic series. In this case, for the functions
$S(x,t)$ and $\varphi_0(x,t)$
we obtain the following problems:
\begin{equation}\label{t.2}
\frac{\partial S}{\partial t}+P\bigg(x,\frac{\partial S}{\partial
x}\bigg)=0,\qquad S(x,t)|_{t=0}=S_0(x),
\end{equation}
\begin{gather}\label{t.3}
\frac{\partial \varphi_0}{\partial t} +\Big(\nabla_{\xi}
P\bigg(x,\frac{\partial S}{\partial x}\bigg), \nabla \varphi_0\Big)
+\frac{1}{2}\sum_{ij}\frac{\partial^2 P}{\partial \xi_i\partial
\xi_j}
\frac{\partial^2 S}{\partial x_i\partial x_j}\varphi_0=0,\\
\varphi_0(x,t)|_{t=0}=\varphi^0(x),\nonumber
\end{gather}

As is known, the solution of problem (\ref{t.2}) is constructed
using the solutions of the Hamiltonian system assumed to exist and
to be smooth
\begin{equation}\label{t.4}
\dot x=\nabla_\xi P(x,p),\qquad x|_{t=0}=x_0,
\end{equation}
$$
\dot p=-\nabla_x P(x,p),\qquad p|_{t=0}=\nabla S_0(x_0).
$$
This solution is smooth on the support of $\varphi_0(x,t)$ for all $t$
such that the
Jacobian $\bigg|Dx/Dx_0\bigg|\ne0$ for
$x_0\in\operatorname{supp}\varphi^0(x)$. We let $g^t_H$ denote the
translation mapping along the trajectories of the Hamiltonian system
(\ref{t.4}).

We recall that the plot
$$
\Lambda^n_0=\{x=x_0,p=\nabla S_0(x_0)\}
$$
is the initial Lagrangian manifold corresponding to Eq.~(\ref{t.2}),
and $\Lambda^n_t=g^t_h\Lambda^n_0$
is the Lagrangian manifold corresponding to
Eq.~(\ref{t.2}) at time $t$.
Let
$\pi:\Lambda^n_t \to \mathbb{R}^n_x$ be the projection
of $\Lambda^n_t$ on $\mathbb{R}^n_x$,
which is assumed to be proper.
For this property to hold, it is sufficient to assume
that the trajectories of the system (2.4)
do not go to infinity during a finite time.

The point $\alpha\in \Lambda^n_t$ is said to be essential if
$$
\hat{S}(\alpha,t)=\min_{\beta\in\pi^{-1}(\alpha)}\hat{S}(\beta,t)
$$
and nonessential otherwise. Here $\hat{S}$ is the action
on $\Lambda^n_t$ determined by the formula
$$
\hat{S}(\beta,t)=\int^t_0 p\,dx-H\,dt,
$$
where the integral is calculated along the trajectories of the
system (\ref{t.4}) the projection of its origin being  $x_0=\beta$.
As is known
$$
S(x,t)=\hat{S}(\pi^{-1}x,t)
$$
at regular points where the projection $\pi$ is bijective.

The global in time asymptotic solution of problem (\ref{t.1}) is
given by the Maslov tunnel canonical operator.

To define this operator, following [8, 10]
we introduce the set of essential points
$\bigcup \gamma_{it}\subset \Lambda^n_t$.
This set is closed because the projection $\pi$ is proper, i.e.
that for all $x$ the set of $p$ such that $(x,p)\in \Lambda^n_t, \pi(x,p)=x$
is finite.

Suppose that the open domains $U_j\subset \Lambda^n_t$ form a
locally finite covering of the set $\bigcup\gamma_{it}$. If the set
$U_j$ consists of regular points, then we set
\begin{equation}\label{t.5}
u_j=e^{-S_j(x,t)/h}\varphi_{0j}(x,t)
\end{equation}
where
$$
\varphi_{0j}(x,t)=\psi_{0j}(x,t)\bigg|\frac{Dx_0}{Dx}\bigg|^{1/2},
$$
$\psi_{0j}(x,t)$ being the solution of the equation
\begin{equation}\label{t.6}
\frac{\partial\psi_{0j}}{\partial t} +( P_{\xi}(x,\nabla
S_j),\nabla\psi_{0j})
-\frac12\operatorname{tr}\frac{\partial^2P}{\partial x\partial\xi}
(x,\nabla S_j)\psi_{0j}=0,
\end{equation}
which exists and is smooth whenever $\bigg|Dx/Dx_0\bigg|\ne0$. The
solution $u_j$ in the domain containing essential (nonregular)
points (at which $d\pi$ is degenerate) is given in the following
way: the canonical change of variables is performed so that the
nonregular points become regular, then we determine a fragment of
the solution in new coordinates by formula (\ref{t.5}) and return to
the old variables, applying the ``quantum'' inverse canonical
transformation to the solution obtained in the new coordinates.

The Hamiltonian determining this canonical transformation
has the form
$$
H_\sigma=\frac12 \sum^{n}_{i=1}\sigma_k p^2_k,
$$
where $\sigma_1,\dots\sigma_n=\operatorname{const}>0$.

The canonical transformation to the new variables is given by the
translation by the time $-1$ along the trajectories of the
Hamiltonian $H_\sigma$. One can prove (see [8],[10]) that the family
of sets $\sigma$ for which the change of variables takes a
nonregular point into a regular is not empty.

Next, the solution near the essential point is determined by the relation
\begin{equation}\label{t.7}
u_j=e^{\frac1h\hat{H}_\sigma}\tilde{u}_j,
\end{equation}
where $\tilde{u}_j$ is given by formula (\ref{t.5}) in the new variables
and
$$
\hat{H}_\sigma=\frac12\sum^{n}_{k=1}\sigma_k
\bigg(-h\frac{\partial}{\partial x_k}\bigg)^2.
$$

On the intersections of singular charts (containing singular points)
and nonsingular charts (without singular points), we must match
$S_j$ and $\psi_{0j}$. This can be done by applying the Laplace
method to the integral whose kernel is a fundamental solution for
the operator $-h\frac{\partial}{\partial t}+\hat{H}_\sigma$. This
integral appears if we write the right-hand side of (\ref{t.7}) in
detail. In this case, since the solution is real, the Maslov index
which is well-known [8] to appear in hyperbolic problems does not
appear. The complete representation of the solution of problem
(\ref{t.1}) is obtained by summing functions of the type (\ref{t.5})
and (\ref{t.7}) over all the domains $U_j$, for more detail, see
[8], [10]. Here we only say the corresponding sum is (locally)
finite because we assumed that the projection $\pi$ is proper.

The asymptotics thus constructed is justified, i.e., the proximity
between the exact and asymptotic solutions of the Cauchy problem
(\ref{t.1}) is proved [8, 9]. More precisely it is proved that at
the  points of the set $\pi(\bigcup\gamma_{it})$ where the
projection $\pi$ is bijective the following estimate holds:
$$
u(x,t,h)-u_j= O(h)
$$

In the preceding Section we noted that the values of the solution of
the continuity equation at nonregular points are independent of the
values of the solution on the singularity support (of course, the
inverse influence takes place) by the condition that the velocity
field trajectories enters the singular support.

In the case of the canonical operator whose construction has briefly
been described above, the relation between the solutions at
essential and nonessential point is also unilateral, namely, the
essential points are ``bypassed'' using (\ref{t.7}), but the values
of the functions
 $u_j$ on the singularity
support do not determine the values at the regular points.

 It is clear that the values of the functions
$S_j$ and $\psi_{0j}$ at regular points are defined by
characteristics via the initial data. The trajectories of the
Hamiltonian system also come to points on $\Lambda^n_t$ such that
the projection mapping at these points is singular. At these points,
we cannot define $\psi_{0j}$ by characteristics directly and we must
use an auxiliary construction (see (2.7)) which allows us to
determine the values of an asymptotic solution at the projections of
singular points onto $x$-space. In this auxiliary construction, the
values of $S_j$ and $\psi_{0j}$ in a neighborhood of the singular
points on $\Lambda^n_t$ are used to determine the values of $u_j$ at
the projections of singular points. Thus, there is a similarity
between this and the preceding sections: we define the values of an
asymptotic solution by characteristics outside its singular support
and then define the values of the asymptotic solution at the
singular points using the already defined values at regular points
in our auxiliary construction.

 Now we note that the function $S(x,t)$  is such that
$$
S(x,t)|_{U_j}=S_j(\pi^{-1}(\alpha),t)
$$
is globally determined and continuous at points of the domain
$\pi(\bigcup\gamma_{it})\subset\mathbb{R}^n_x$. We denote this set
by $\bigcup\Gamma_{i}$ and assume that this is a stratified
manifold with smooth strata $\Gamma_{it}$ of different codimensions.
We note that, for example, if the inequality
$\nabla(S_i(x,t)-S_j(x,t))\ne0$ holds while we pass from one branch
$\Lambda^n_t\cap\bigcup\gamma_{it}$ to another, then the set
$\pi\{(\tilde{S}_i-\tilde{S}_j)=0\}$ generates a smooth stratum of
codimension $1$. In the one-dimensional case, all strata are points
or curves on the $(x,t)$-plane (under the above assumptions about the
singularities being discrete).

Now we consider the equation for $\psi^2_{0j}$.
We denote this function by $\rho$ and then obtain
\begin{equation}\label{t.8}
\frac{\partial \rho}{\partial t}
+(\nabla,u\rho)+a\rho=0,
\end{equation}
where $u(x,t)=\nabla_{\xi} P(x,\nabla S)$ and $a=-\operatorname{tr}
\frac{\partial^2 P}{\partial x\partial\xi}(x,\nabla S)$.

If the condition
$$
\operatorname{Hess_{\xi}} P(x,\xi)>0
$$
is satisfied, then it follows from the implicit function theorem
that $\nabla S(x,t)=F(x,u(x,t))$, where $F(x,u)$ is a smooth
function and
$$
a=f(x,u),
$$
where $f(x,z)$ is again a smooth function.

Thus, we have proved the following theorem.

\begin{theorem}
Suppose that the following conditions are satisfied
for $t\in[0,T]$, $T>0$:

{\rm(1)} There exists a smooth solution of the Hamiltonian system
{\rm(\ref{t.4})}.

{\rm(2)} The singularities of the velocity field $u=\nabla_\xi
P(x,\nabla S)$ form a stratified manifold with smooth strata and
$\operatorname{Hess_\xi} P(x,\xi)>0$.

{\rm(3)} There exists a generalized solution $\rho$ of the Cauchy problem
for Eq.~{\rm(\ref{t.8})} in the sense of the integral identity {\rm(\ref{2.7})}.

Then at the points of $\pi(\bigcup\gamma_{it})$ where the projection
$\pi$ is bijective, the asymptotic solution of the Cauchy problem
{\rm(\ref{t.1})} has the form
$$
u=\exp(-S(x,t)/h)(\sqrt{\rho}+O(h)).
$$
\end{theorem}

This theorem is a global in time analog of the corresponding
Ma\-de\-lung observation [13] about local solutions of Schroedinger
type equations.

\section{Particular cases}

  The theorem stated in the previous section requires that some assumptions
are satisfied. The most restrictive is the item 3 in theorem 2, that
is the existence of the global generalized solution to the
continuity equation. Under the above made assumptions it is possible
to construct this solution using characteristics, but only in the
case where the structure of the singular support of $u$ is not
changing in time-all sections of the stratified manifold introduced
above by planes $t=\text{const}$  smoothly depend on $t$. A more
complicate situation arises when the singularities of the velocity
field change their structure. In this case the problem of the
construction of a global in time generalized solution to the
continuity equation has not been solved yet. The obstacle is that in
this case usually one has no global in time expression for the
velocity field $u$. In turn this does not allow to apply formula
(\ref{3.8}) to construct global solution to the continuity equation.
In the multy-dimensional case as far as we know there is only one
result concerning to shock wave generation [14] which allows to
construct global in time approximations of the shock wave formation
process. But this is slight different from the construction that we
needs here. In the one dimensional case the situation is better and
we have all needed formulas.

We begin with the spatially homogeneous case. Here the problem is
equivalent to the one of constructing a formula for a global
solution to conservation law equation

\begin{equation}\label{d.1}
\frac{\partial v}{\partial t}+\frac{\partial P(v,t)}{\partial x}=0
\end{equation}

Here $P(-h\frac{\partial}{\partial x},t)$ is the same operator as in
(\ref{t.1}) but assumed to be independent of x with the symbol $P(\xi,t)$ and
$v={\partial S}/{\partial x}$. The velocity field $u$  in this case
is $P_{\xi}(v,t)$. In [11] a construction of the
global solution to the continuity equation where the velocity field
is given by the solution of the equation (\ref{d.1}) was given. Because the
set of singular points is discrete by our assumptions, without loss
of generality one can consider the case were only one point of
singularity appears. Denote the corresponding (smooth) initial
condition by $u_0$, the instant where the singularity appears by $t^*$
and the point of singularity by $x^*$.

\textbf{The first step} of construction suggested in [11], [12] is
that we change $u_0$ in a small neighborhood of of origin $x_0^*$ of
the trajectory coming to $x^*$ when $t=t^*$. We denote this new part
of initial data $u_1(x_0)$ for $x_0\in(x_0^*-\beta, x_0^*+\beta)$,
$\beta\to0$ and assume

\begin{equation}\label{d.5}
\varepsilon\beta^{-1}\to0, \qquad \varepsilon\to0.
\end{equation}

We define the function $u_1=u_1(x_0,t)$ as a solution of implicit
equation
\begin{equation}\label{d.11}
P'_\xi(u_1,t)=- K(t)x_0+b(t),
\end{equation}
The latter equation is solvable under the condition
$\operatorname{Hess_\xi} P(x,\xi)>0$, formulated above.

The functions $K(t)$ and $b(t)$ are defined from the condition of continuity
of the characteristics flow, i.e
$$
u_1(x_0^*-\beta,t)=u_0(x_0^*-\beta,t),\qquad
u_1(x_0^*+\beta,t)=u_0(x_0^*+\beta,t)
$$

 It is easy
to check that this choice of $u_1$ provides that the Jacobian
$\big|Dx/Dx_0\big|$ is identically equal to $0$ for $t=t^*$ and
$x_0\in(x_0^*-\beta, x_0^*+\beta)$. Here we remove from usual
topological concept of general position considering the situation of
identical equality that can be destroyed by small perturbation. But
this construction follows from the algebraic concept  and allows to
present the solution of (\ref{d.1}) in the form of linear
combination of Heaviside functions (see [11]).

\textbf{The second step} of our construction of an approximation is a
modification of the definition of characteristics. We set
\begin{equation}\label{d.2}
\dot x=(1-B)P'_{\xi}(u_1(x_0,t ),t)+Bc,\qquad
x_0\in(x_0^*-\beta,x_0^*+\beta)
\end{equation}

and

$$
\dot x=P'_{\xi}(u_0,t),
$$

where $x_0$ does not belong to  $(x_0^*-\beta,x_0^*+\beta)$,
$$
c=\frac{P(v(x(x_0^*+\beta,t),t))-P(v(x(x_0^*-\beta,t),t))}{v(x(x_0^*+\beta,t),t)-v(x(x_0^*-\beta,t),t)}
$$
The initial data for (\ref{d.2}) are as follows:
$$
x\bigg|_{t=0}=x_0+A\varepsilon,\qquad \varepsilon>0.
$$
The function B in (\ref{d.2}) has the form
$B=B((t-t^*)/\varepsilon)$ and B(z) is smooth, monotone and
increasing from 0 to 1 for $z\in(-\infty,\infty)$. Similarly to
[11], [12] one can prove that there exist an
$A=\operatorname{const}$ such that the Jacobian $\big|Dx/Dx_0\big|$
calculated using the above introduced characteristics is not equal
to zero, but it is of order $O(\varepsilon)$ when
$t\ge{t^*+O(\beta)}$ when $x_0\in(x_0^*-\beta,x_0^*+\beta)$. Using
the velocity field generated by $\dot x$ we can construct global in
time (smooth) solution of the continuity equation in the form
(\ref{3.8}). After that, passing to the limit as $\varepsilon\to0$
we will obtain the generalized solution of the continuity equation in
the sense of definition from Sec.1 just like it was done in [12].

\textbf{Spatially inhomogeneous one dimensional case}.

We will follow the scheme introduced above. The case under
consideration can be treated in the same way as the previous one
with modifications. Firstly, we will assume that the symbol
$P=P(x,\xi)$ does not depend on t. In this case this assumption
(which means that the mapping $g^t_P$ is invertible w.r.t. time)
will be used to construct the insertion to initial data. In the
previous case we did it using the implicit function theorem, see
(\ref{d.11}).

Let $\Lambda^1_0$ be a smooth nonsingular (w.r.t the projection
$\pi$) curve in the $(x,p)$ space, which is a Lagrangian manifold
corresponding to initial data for our problem. We consider the
Lagrangian manifold $\Lambda^1_{t^*}=g^{t^*}_P\Lambda^1_0$  and
assume that there is only one point singular with respect to the
projection onto $x$-axis and its projection is $x^*$. Let $\beta$ be
the same as above. Let us set $t_1^*=t^*+\beta$. Because of the
assumption that $P''_{\xi\xi}$ is positive, we have that for
$t=t_1^*$ the Lagrangian manifold $\Lambda^1_{t^*_1}$ has two parts
which contain essential points and these parts form a shock wave
type curve with the jump at the point $x^*_1$ where
$S_{left}(x^*_1,t^*_1)=S_{right}(x^*_1,t^*_1)$. We connect these
parts by a vertical line and thus obtain a new Lagrangian manifold,
which is a piecewise smooth continuous curve with two angle points
(ends of the vertical part, the distance between them of order
$\beta$). We denote this manifold by $\hat\Lambda^1_{t^*_1}$ and
apply the mapping $g_P^{-t_1}$ for sufficiently small $t_1$ to this
manifold. This mapping obviously exists  and is a diffeomorpfism
because our Hamiltonian $P$ does not depend on $t$. We consider the
obtained manifold $g_P^{-t_1}\hat\Lambda^1_{t^*_1}$ as the new
Lagrangian manifold corresponding to our problem for
$t=t^*_1-t_1$changing the manifold $\Lambda^1_{t^*_1-t_1}$ by
$g_P^{-t_1}\hat\Lambda^1_{t^*_1}$. As it was said above the latter
manifold is piecewise smooth curve with two angle points and all
points of the curve outside of the part between these angle points
are regular. Moreover there exist a sufficiently small $t_1$ such
that the part of the curve between these angle points contains only
regular points-these statements are a consequence of the positivity
of $P''$,  its stationarity and the possibility to choose $t_1$
small enough (and independent on $\varepsilon$).

Denote the projections of the mentioned above angle points on the
manifold $g_P^{-t_1}\hat\Lambda^1_{t^*_1}$ to the $x$-axis by $a_1<
a_2$. We note that $\big|a_1-a_2\big|$ is of order $\beta$.

Like in the previous example we introduce the new characteristics
system
\begin{equation}\label{d.3}
\dot x=(1-B)P'_{\xi}(x(x_0,t ),p(x_0,t))+Bc,
\end{equation}
$$
\dot p=-(1-B)P'_x(x(x_0,t ),p(x_0,t),\qquad x_0\in(a_1,a_2),
$$

and

\begin{equation}\label{d.7}
\dot x=P'_{\xi}(x,p),
\end{equation}
$$
\dot p=-P'_x(x,p)
$$
when $x_0$ does not belong to  $(a_1,a_2)$. We have set
\begin{equation}\label{d.8}
c=\frac{P(v(x(a_2,t),p(a_2,t)))-P(x(a_1,t),p(a_1,t))}{p(x(a_2,t),t)-p(x(a_1,t),t)}
\end{equation}
The initial data for (\ref{d.3}), (\ref{d.7}) are as follows:

$$
x \big|_{t=0}=x_0+A\varepsilon,
$$
$$
p \big|_{t=0}=p_0(x_0),
$$
where $(x_0,p_0(x_0))$=$g_P^{-t_1}\hat\Lambda^1_{t^*_1}$. The
expression on the right hand side of (\ref{d.8}) is the direct
analog of the well known Rankine-Hugoniot expression for the
velocity of the shock propagation. In the case under consideration
it is the velocity of the point $\check x$ on $x$-axis, where
$S_{left}(\check x,t)=S_{right}(\check x,t)$.

By the assumption we have only one singular point  if we are
considering the family of manifolds $\Lambda^1_t$, $t\in[0,t^*]$. We
also have by construction that the Jacobian $J= Dx/Dx_0$ calculated
using  the solutions of the system (\ref{d.3}) is not equal to zero.
More precisely we have
$$
\lim_{\varepsilon\to0} J=H(t^*-t)J_0,
$$
where $J_0$ is the Jacobian calculated using the solutions of
(\ref{d.3}) for $B=0$ ($J_0=0$ when $t=t^*$ by construction) and
$$
J \ge H(t^*-t)J_0+C\varepsilon,
$$
where $C=\operatorname{const}>0$. This statement directly follows
from (\ref{d.3}) if we take the properties of the function $B$ into
account. This means that the velocity field, generated by the
projections of the solution of the system (\ref{d.3}), (\ref{d.7})
onto the $x$-axis has nonintersecting trajectories for
$\varepsilon>0$. Thus we can use it to construct solutions of the
continuity equation. It remains to note that just like in [12] it is
easy to check that the limits of these solutions will satisfy to the
integral identities introduced in Sec. 1 as the definition of
generalized solutions to continuity equation.

\end{document}